\documentclass[aps,twocolumn,pra,superscriptaddress]{revtex4-1}

\usepackage[english]{babel}
\usepackage{amsfonts}
\usepackage{amssymb}
\usepackage{amsmath}
\usepackage{graphicx}
\usepackage{mathtools}

\usepackage{color}

\newcommand{\expo}[1]{\mathrm{exp}\left(#1\right)}
\newcommand{\cose}[1]{\mathrm{cos}\left(#1\right)}
\newcommand{\seno}[1]{\mathrm{sin}\left(#1\right)}
\newcommand{\ex}[1]{e^{#1}}
\newcommand{\Ket}[1]{\left|#1\right>}
\newcommand{\Bra}[1]{\left<#1\right|}
\newcommand{\BraKet}[2]{\left<#1|#2\right>}
\newcommand{\KetBra}[2]{|#1\rangle\langle#2|}

\newcommand{\conm}[2]{\left[#1,#2\right]}
\newcommand{\etal}{\emph{et al.} }
\newcommand{\eps}{\varepsilon}

\begin{document}

\title{Analysis of lower bounds for quantum control times and their relation to the quantum speed limit \\
\vspace{5pt}
An\'alisis de cotas inferiores para tiempos de control y su relaci\'on con el l\'imite de velocidades cu\'antico}

\author{Pablo M. Poggi}
\email{Corresponding author: ppoggi@unm.edu}
\affiliation{Center for Quantum Information and Control (CQuIC), Department of Physics and Astronomy, University of New Mexico, Albuquerque, New Mexico 87131, USA}
\affiliation{Departamento de F\'isica ``J. J. Giambiagi'' and IFIBA, FCEyN, Universidad de Buenos Aires, 1428, Buenos Aires, Argentina}

\date{\today}

\begin{abstract}
Limitations to the speed of evolution of quantum systems, typically referred to as quantum speed limits (QSLs), have important consequences for quantum control problems. However, in its standard formulation, is not straightforward to obtain meaningful QSL bounds for time-dependent Hamiltonians with unknown control parameters. In this paper we present a short introductory overview of quantum speed limit for unitary dynamics and its connection to quantum control. We then analyze potential methods for obtaining new bounds on control times inspired by the QSL. We finally extend the work in [Poggi, Lombardo and Wisniacki \textbf{EPL} 104 40005 (2013)] by studying the properties and limitations of these new bounds in the context of a driven two-level quantum system.\\

Las restricciones a la velocidad de evoluci\'on de un estado cu\'antico, usualmente llamadas ``l\'imite de velocidades cu\'antico'' (QSL),  es un concepto que presenta importantes consecuencias para problemas de control cu\'antico. Sin embargo, en su formulaci\'on usual, no es trivial obtener cotas inferiores tipo QSL para el tiempo de evoluci\'on en el caso de Hamiltonianos dependientes del tiempo con par\'ametros desconocidos. En este trabajo presentamos un introducci\'on a la formulaci\'on del l\'imite de velocidades cu\'antico para evoluci\'on unitaria y su conexi\'on con control cu\'antico. Luego, analizamos nuevos m\'etodos para obtener cotas inspiradas en el QSL para tiempos de evoluci\'on en problemas de control. Finalmente, extendemos el trabajo presentado en [Poggi, Lombardo and Wisniacki \textbf{EPL} 104 40005 (2013)] estudiando las propiedades y limitaciones de las cotas presentadas en un sistema de dos niveles.

\end{abstract}

\maketitle

\section{Introduction}

Precise control of the dynamics of microscopic systems is a cornerstone of the ongoing revolution in quantum technologies like quantum computation and simulation. Indeed, most physical implementations of quantum devices rely on accurate and robust manipulation of the relevant degrees of freedom using time-dependent electromagnetic fields \cite{Bloch2012,zhang2017_monroe,bernien2017}. Such advances where made possible by substantial technological breakthroughs but also by theoretical developments in the field of quantum control \cite{dalessandro,glaser2015}. A crucial part of this theory is related to implementing the desired transformations on a quantum system as fast as possible, in order to avoid undesirable environmental effects which can destroy the coherence properties of the system \cite{schlosshauer2007}. In this context, during the past two decades there has been a renewed interest on understanding the fundamental limitations on the speed of evolution of quantum systems. These limitations, typically referred to as quantum speed limits (QSLs), were originally formulated via Heisenberg-like uncertainty relations by Mandelstam and Tamm in the mid 20th century \cite{mand_tamm1945}, and have since then been thoroughly studied and generalized to a variety of scenarios, such as open quantum system dynamics, evolution of mixed states and time-dependent Hamiltonians \cite{fleming1973,bhatta1983,giovannetti2003,anandan1990,taddei2013,deffner2013,delcampo2013,deffner2013_teur}. \\

The connection between the QSL and practical quantum control problems received much attention since the work of Caneva \etal \cite{caneva2009}, who showed that quantum optimal control methods \cite{krotov1999} could be used to explore what is the minimal time needed to control a quantum system, and provided a link with the QSL \footnote{The nomenclature can be confusing since the quantum control literature typically refers to minimum control times as 'quantum speed limit times'. Such quantity is not directly related to the original quantum speed limit results given by the Mandelstam-Tamm (and also Margolus-Levitin). The main difference is that the minimum control time depends on a target state, while the QSL time does not} bounds for some specific systems. Since then, numerous studies have implemented this methodology \cite{caneva2011,tibbetts2012,brouzos2015,poggi2015_1,sorensen2016,arenz2017}. However, apart from a handful of cases \cite{khaneja2002,boozer2012,albertini2015}, the search for the minimum control time has to be performed numerically and, even in that case, one can only find an upper bound to it \cite{sorensen2016}. So, as has been pointed out in previous works \cite{arenz2017,poggi2019}, it is important to develop lower bounds on control times which are as informative and tight as possible, while at the same time being computable before solving the actual (optimal) control problem. In this paper we illustrate how the standard QSL formulation is not particularly suitable for this task, because of its dependence on the (\textit{a priori} unknown) evolution on the system. To demonstrate this point, we present a self-contained introduction to the standard QSL formulation for unitary dynamics and its application to time-dependent Hamiltonians. We then show that the presented framework, suitable extended and modified, can indeed lead to meaningful lower bounds on the control time. We show three examples of such bounds which are taken or adapted from previous works, and explicitly work them out for the paradigmatic example of state control on a driven two-level quantum system. \\

This paper is organized as follows. In Sec. \ref{sec:qsl_overview} we present an introductory overview on the topic of quantum speed limits for unitary evolution, going through its original formulation as derived from Robertson's uncertainty relation, and its geometrical interpretation due to Anandan and Aharonov. Then, in Sec. \ref{sec:qcontrol} we discuss QSLs for time-dependent Hamiltonians and its corresponding natural connection with quantum control. Here we argue that the QSL bounds derived in this formulation cannot generally be used for bounding control times \textit{a priori}, i.e., before solving the optimal control problem, because of the presence of unknown control parameters. We then revisit scattered proposals in the literature of bounds which overcome this issue and discuss their connection with the standard QSL. Finally, in Sec. \ref{sec:two_level} we compare the aforementioned bounds in the context of a driven two-level system. In this way we extend the results of Ref. \cite{poggi2013}, in which different bounds derived from the standard QSL where compared originally. At the end of the article, in Sec. \ref{sec:final} we present some ideas for future work and final remarks.

\section{Quantum speed limit formulation for unitary evolution}
\label{sec:qsl_overview}

Here we present an introductory overview of the quantum speed limit formulation for Hamiltonian evolution, including derivations of the most relevant mathematical expressions. Note that we do not discuss extensions and generalizations beyond unitary dynamics; the reader interested in a complete review on this topic is advised to consult Ref. \cite{deffner2017}.

\subsection{Overview}

In 1945, Mandelstamm and Tamm \cite{mand_tamm1945} derived a generalization of Heisenberg uncertainty relation between time and energy, that could be applied to any quantum system. We re-derive it here, starting from Robertson's inequality \cite{robertson1930}
\begin{equation}
\langle(\delta A)^2\rangle\langle(\delta B)^2\rangle\geq\frac{1}{4}\left|\langle\conm{A}{B}\rangle\right|^2,
\label{ec:qsl_robert}
\end{equation}
\noindent where $\delta A=A-\langle A\rangle$. For any operator $A$ we can write Heisenberg's equation
\begin{equation}
\frac{dA}{dt}=-\frac{i}{\hbar}\conm{A}{H}.
\end{equation}
By taking the expectation value in the last expression we obtain
\begin{equation}
\frac{d\langle A\rangle}{dT}=-\frac{i}{\hbar}\langle\conm{A}{H}\rangle.
\label{ec:qsl_ehren}
\end{equation}
We now identify operator $B$ in eqn. (\ref{ec:qsl_robert}) with the system Hamiltonian $H$ and combine with eqn. (\ref{ec:qsl_ehren}) to obtain
\begin{equation}
\Delta E \Delta A\geq\frac{\hbar}{2}\left|\frac{d\langle A\rangle}{dt}\right|,
\label{ec:qsl_mt1}
\end{equation}
\noindent where $\Delta A=\sqrt{\langle A-\langle A\rangle\rangle^2}$, and $\Delta E\equiv \Delta H$. We can further define
\begin{equation}
\Delta t_{A}=\frac{\Delta A}{\left|\frac{d\langle A\rangle}{dt}\right|},
\label{ec:qsl_tloco}
\end{equation}
\noindent which has units of time. We then arrive at the Mandelstam-Tamm relation 
\begin{equation}
\Delta t_A \Delta E \geq \frac{\hbar}{2}.
\label{ec:qsl_mt}
\end{equation}
In this formulation, $\Delta t_A$ is interpreted as a characteristic time related to the time evolution of observable $A$. The link between this quantity and the physical evolution time was studied first by Fleming \cite{fleming1973} and then by Bhattacharyya \cite{bhatta1983}, in the following way. Consider expression (\ref{ec:qsl_mt}) under the specific choice of $A=\KetBra{\phi_0}{\phi_0}$, with $\Ket{\phi_0}$ some arbitrary pure state. If we take the expectation values in (\ref{ec:qsl_mt}) with respect to the evolved state $\Ket{\phi_t}=U_t\Ket{\phi_0}$, it is easy to see that
\begin{equation}
\langle A\rangle=\lvert\BraKet{\phi_t}{\phi_0}\rvert^2=P_t,
\end{equation}

\noindent where we have introduced the short-hand notation for $P_t$, the time-dependent survival probability. Eqn. (\ref{ec:qsl_mt}) can now be expressed as
\begin{equation}
\frac{\left|\frac{dP_t}{dt}\right|}{\sqrt{P_t(1-P_t)}}\leq 2\frac{\Delta E}{\hbar}.
\label{ec:qsl_mt2}
\end{equation}

We can use the relation $\frac{d}{dx}\left[\mathrm{arccos}(x)\right]=-(1-x^2)^{-1/2}$ to write (\ref{ec:qsl_mt2}) in a more compact form
\begin{equation}
\frac{d}{dt}\mathrm{arccos}(\sqrt{P_t})\leq\frac{\Delta E(t)}{\hbar}.
\label{ec:qsl_bhatta}
\end{equation} 

This is the main result by Bhattacharyya. If the initial state $\Ket{\phi_0}$ evolves subject to a time-independent Hamiltonian $H$, then the inequality above can be readily integrated from $t=0$ to $t$, obtaining
\begin{equation}
t \geq\frac{\hbar}{\Delta E}\mathrm{arccos}\left(\left|\BraKet{\phi_0}{\phi_t}\right|\right)\equiv t_{QSL}^{MT}.
\label{ec:qsl_mtbound}
\end{equation} 

This is the Mandelstam-Tamm bound. In the particular case where $\Ket{\phi_t}$ is orthogonal to $\Ket{\phi_0}$, we obtain $t_{QSL}=\frac{\pi \hbar}{2\Delta E}$. This expression sets a bound on the minimum time required for a system to evolve from $\Ket{\phi_0}$ to an orthogonal state. For completeness we mention that, for this case, Margolus and Levitin \cite{margolus1998} also derived a similar bound, but in terms of the mean energy of the state,
\begin{equation}
t\geq \frac{\pi\hbar}{2\:E}\equiv t_{QSL}^{ML},
\label{ec:qsl_ml}
\end{equation}

\noindent where $E\equiv\langle H-\varepsilon_0\mathbb{I}\rangle$, i.e. the expectation value of the Hamiltonian with respect to the ground state. Giovannetti \etal \cite{giovannetti2003} later generalized this result to non-orthogonal states, and coined the term ``quantum speed limit time'' for $t_{QSL}$. Finally, Levitin and Toffoli \cite{levitin2009} showed that the unified bound
\begin{equation}
t\geq\mathrm{min}\left\{\frac{\pi\hbar}{2\:\Delta E},\frac{\pi\hbar}{2\:E}\right\},
\label{ec:qsl_unif}
\end{equation}

\noindent is tight, meaning that for every time-independent Hamiltonian there is a choice of initial state for which the equality in (\ref{ec:qsl_unif}) holds.

\subsection{Geometric quantum speed limits}

\begin{figure}[t]
	\centering
	\includegraphics[width=.8\linewidth]{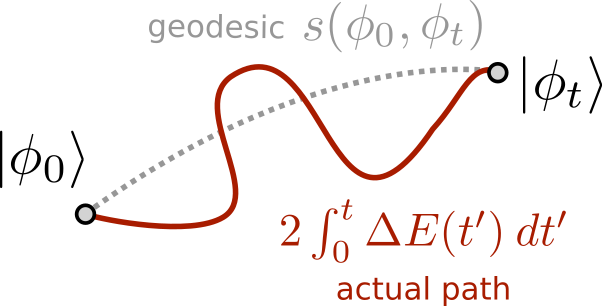}
	\caption{Schematic drawing of the time evolution of quantum states. Anandan-Aharonov relation (\ref{ec:qsl_aa}) expresses the fact that the length of the actual path of the evolution is necessarily larger or equal than the length of the geodesic path between the initial and evolved state.}
	\label{fig:qsl_geod}
\end{figure}

Bhattacharyya's result of eqn. (\ref{ec:qsl_bhatta}) has an insightful geometrical interpretation, which was first noted by Anandan and Aharonov \cite{anandan1990} in the following way. Consider the Fubini-Study distance between two pure states,
\begin{equation}
s(\phi_1,\phi_2)=2\:\mathrm{arccos}(\left|\BraKet{\phi_1}{\phi_2}\right|),
\label{ec:qsl_fs}
\end{equation}


\noindent and define $ds=s\left(\phi_t,\phi_{t+dt}\right)$ with
\begin{equation}
\Ket{\phi_{t+dt}}=\ex{-\frac{i}{\hbar} H(t)dt}\Ket{\phi_t}
\end{equation}

\noindent for some state $\Ket{\phi_t}$ and a generally time-dependent Hamiltonian $H(t)$. Since 
\begin{equation}
\left|\BraKet{\phi_t}{\phi_{t+dt}}\right|^2 = 1-\frac{1}{\hbar^2}\Delta E(t)^2\: dt^2+\mathcal{O}(dt^4),
\end{equation}

\noindent then the differential length element is given by
\begin{equation}
ds = \frac{2}{\hbar}\Delta E(t) dt
\label{ec:qsl_aa_dif}
\end{equation}

\noindent which is formally eqn. (\ref{ec:qsl_bhatta}) rewritten with different notation. Integration of eqn. (\ref{ec:qsl_aa_dif}) from $t=0$ to $t$ yields the length of the path traversed by the evolution going from the initial state $\Ket{\phi_0}$ to the evolved state $\Ket{\phi_t}$. Clearly, such length must be greater or equal than $s(\phi_0,\phi_t)$, the length of the geodesic path joining both states. This can be appreciated in the schematic drawing of Fig. \ref{fig:qsl_geod}. Thus, we have derived the Anandan-Aharonov relation 

\begin{equation}
s(\phi_0,\phi_t)\leq 2\int_0^t \Delta E(t')\:dt',
\label{ec:qsl_aa}
\end{equation}

\noindent where we have (finally) set $\hbar=1$. \\

Note that expression (\ref{ec:qsl_aa_dif}) also tells us that energy variance $\Delta E(t)$ can be seen as a measure of the Hilbert space velocity of the state $\Ket{\phi_t}$. In particular, $\Delta E$ measures the component of $\dot{\Ket{\phi_t}}$ which is perpendicular to $\Ket{\phi_t}$ \cite{pati1995,carlini2006,gajdacz2015}. We can see this in the following way. If we write the time derivative of the quantum state as $\dot{\Ket{\phi_t}}=\dot{\Ket{\phi_t}}^{\parallel}+\dot{\Ket{\phi_t}}^{\perp}$, then we have that, by definition,
\begin{equation}
\dot{\Ket{\phi_t}}^\parallel=\Ket{\phi_t}\BraKet{\phi_t}{\dot{\phi_t}}=-i\langle E\rangle\Ket{\phi_t},
\end{equation}

\noindent where we have used $\dot{\Ket{\phi_t}}=-i H_t\Ket{\phi_t}$ and noted $\Bra{\phi_t}H_t\Ket{\phi_t}\equiv\langle E\rangle$. This result tells us that the phase of the quantum state evolves at a rate given by $\langle E\rangle$. The remaining perpendicular component of the velocity, $\dot{\Ket{\phi_t}}^\perp=\dot{\Ket{\phi_t}}-\dot{\Ket{\phi_t}}^\parallel$, is such that
\begin{eqnarray}
\|\dot{\Ket{\phi_t}}^\perp\|^2&=&\BraKet{\dot{\phi_t}}{\dot{\phi_t}}+\BraKet{\dot{\phi}_t^\parallel}{\dot{\phi}_t^\parallel}-\BraKet{\dot{\phi}_t}{\dot{\phi}_t^\parallel}-\BraKet{\dot{\phi}_t}{\dot{\phi}_t^\parallel} \nonumber \\
&=& \langle H^2 \rangle-\langle H\rangle^2=\Delta E^2.
\end{eqnarray}

It can be readily seen that the Mandelstam-Tamm bound is recovered from the Anandan-Aharonov relation when the dynamics is generated by a time-independent Hamiltonian, in which $\Delta E$ is always time-independent itself. As such, the inequality (\ref{ec:qsl_mtbound}) has a purely geometrical nature, and its saturated if and only if the motion of the system state is along a geodesic in Hilbert space.\\

\subsection{Extensions and other studies}

Most of the extensions and generalizations of the quantum speed limit formulation have been pursued in this geometrical setting. In particular, bounds have been derived for the maximum speed of evolution under non-unitary dynamics almost simultaneously by Taddei \etal \cite{taddei2013}, Del Campo \etal \cite{delcampo2013} and Deffner and Lutz \cite{deffner2013}. Special attention has been devoted to studying the predicted speed-up of the evolution in open systems undergoing non-Markovian dynamics \cite{cimmarusti2015,sun2015,mirkin2016,cianciaruso2017}. Other important cases of study are QSLs for mixed states \cite{andersson2014,zhang2014,mondal2016,mondal2016_2,marvian2016,campaioli2018}, the geometric characterization of the QSL \cite{russell2014,pires2016,deffner2017_geom,campaioli2019} and its connection to parameter estimation theory \cite{taddei2013,pang2014,gessner2018,sidhu2020}. Extensive analysis of the current state of knowledge on these topics have been published as reviews in Refs. \cite{frey2016,deffner2017}.\\

\section{Connection to quantum control}
\label{sec:qcontrol}
\subsection{QSL for time-dependent Hamiltonians}

Consider a quantum system initially prepared in state $\Ket{\psi_0}$, which evolves according to a Hamiltonian $H(\vec{u}(t))$, where $\vec{u}(t)$ is a set of generally time-dependent parameters (the control fields). We wish to drive the system to some target state $\Ket{\psi_g}$ at some final time $T$ by properly choosing $\vec{u}(t)$. It is natural to ask then, what does the quantum speed limit formulation tells us about the time $T$ required to perform that process? Can it be made arbitrarily fast? Can we establish a lower bound for $T$?\\

At first glance, it is obvious that nor the Mandelstam-Tamm (\ref{ec:qsl_mt2}) nor the Margolus-Levitin (\ref{ec:qsl_ml}) bounds can be applied to this setting, since quantum control problems deal generally with time-dependent Hamiltonians. We then go back to the Anandan - Aharonov relation (\ref{ec:qsl_aa}) to obtain a bound on the evolution time. This can be done in a number of ways: one of them was proposed by Deffner and Lutz \cite{deffner2013_teur}, and it simply consists on rewriting eqn. (\ref{ec:qsl_aa}) as
\begin{equation}
t\geq\frac{\mathrm{arccos}\left(\left|\BraKet{\psi_0}{\psi(t)}\right|\right)}{\overline{\Delta E}},
\label{ec:qsl_deff1}
\end{equation}

\noindent where we defined the time-average of the energy variance simply as
\begin{equation}
\overline{\Delta E}=\frac{1}{t}\int_0^t\: \Delta E (t')\:dt'.
\end{equation}

We can now evaluate (\ref{ec:qsl_deff1}) in $t=T$, such that if there is a time $T$ such that $\Ket{\psi(T)}=\Ket{\psi_g}$, then the following relation must hold
\begin{equation}
T\geq\frac{\mathrm{arccos}\left(\left|\BraKet{\psi_0}{\psi_g}\right|\right)}{\overline{\Delta E}}\equiv T_{QSL}^*.
\label{ec:qsl_deff2}
\end{equation}

However, a closer look at expression (\ref{ec:qsl_deff2}) reveals that, in order to compute the bound, we need both an actual choice of $u(t)$ and the complete time-evolved state $\Ket{\psi(t)}$. This contradicts our initial purpose, which is to estimate the minimum evolution time without solving the dynamics, and moreover without knowing the actual control field which will be used to drive the system. Further insight can be obtained by casting the expression (\ref{ec:qsl_deff2}) into the form
\begin{equation}
T_{QSL}^*=\frac{s(\psi_0,\psi_g)}{\int_0^T \Delta E(t')\: dt'}T=\frac{s_{\mathrm{geod}}}{s_{\mathrm{path}}}T.
\end{equation}

In the last expression, we can see that the lower $T_{QSL}^*$ depends on two geometrical quantities: the length of the geodesic between $\Ket{\psi_0}$ and $\Ket{\psi_g}$ and the length of the actual path. Moreover, the quantum speed limit time could go to zero if the $s_{\mathrm{path}}\gg s_{\mathrm{geod}}$. It is then clear that this quantity gives us information about distances in Hilbert space, but not about the speed at which those paths are traversed. We also point out that other bounds on the evolution time can be extracted from the general Anandan - Aharonov relation (see \cite{mirkin2016} for an example). However, as discussed in Ref. \cite{poggi2013}, in all cases information about the evolution of the system is required to compute such bounds. \\


\subsection{Methods for bounding control times}

In the previous subsection we showed that the usual quantum speed limit formulation is in general not suitable for obtaining bounds on the evolution time of a controlled quantum system \emph{a priori} (i.e., without needing to solve the Schr\"{o}dinger equation). Here, we analyze various methods to overcome this limitation.\\

We begin by explicitly formulating the problem of interest. Consider a quantum system which evolves unitarily under the action of a parameter-dependent Hamiltonian $H(\vec{u})$, with $\vec{u}=\vec{u}(t)$ the (generally time-dependent) control fields. Although the form of the time-dependence is unknown \textit{a priori}, we consider that the control fields may have constraints of the form $|u_i(t)|\leq u_i^{max}$. Let us fix an initial state $\Ket{\psi_0}$ and a target state $\Ket{\psi_g}$. We wish to obtain a lower bound on the evolution time $T$, where $T$ is such that $\Ket{\psi(0)}=\Ket{\psi_0}$ and $\Ket{\psi(T)}=\Ket{\psi_g}$. The bound should be computable with all given information, i.e., it should be of the form
\begin{equation}
T\geq t_{min}\left(H,\{u_i^{max}\},\Ket{\psi_0},\Ket{\psi_g}\right).
\label{ec:qsl_new_gen}
\end{equation}

Our first approach to this problem is to manipulate the Anandan - Aharonov relation (\ref{ec:qsl_aa}) in order to drop any implicit or explicit dependence on $\Ket{\psi(t)}$ or $\vec{u}(t)$. This can be done by using the following inequality
\begin{equation}
2 \Delta E(t) \leq \sqrt{2}\|H(t)\|\equiv\sqrt{2\:\mathrm{tr}(H(t)^2)},
\label{ec:qsl_brody}
\end{equation}

\noindent which was derived by Brody in \cite{brody2015}. 
Combining (\ref{ec:qsl_aa}) and (\ref{ec:qsl_brody}) we can write
\begin{equation}
s(\psi_0,\psi(T))\leq \sqrt{2} \int_0^T \|H(t')\|\:dt'\leq \sqrt{2} \|H\|_{max}T.
\label{ec:qsl_new_brody2}
\end{equation}

In the last step, we bounded $\|H\|$ by its maximum value, which will be a function of $\{u_i^{max}\}$ in general. In this way we have successfully derived an inequality without using information about $\Ket{\psi(t)}$ nor $\vec{u}(t)$. Rearranging the last expression, we obtain that if there is a time $T$ for which $\Ket{\psi(T)}=\Ket{\psi_g}$, then it holds that

\begin{equation}
T\geq\frac{s(\psi_0,\psi_g)}{\sqrt{2}\|H\|_{max}}\equiv t_{min}^{A}.
\label{ec:qsl_new_brody}
\end{equation}

Note that the definition of $t_{min}^A$ is clearly of the form we initially proposed, c.f. eqn. (\ref{ec:qsl_new_gen}).\\

Another approach to obtain a bound of the form (\ref{ec:qsl_new_gen}) can be derived from a result by Pfeifer in Refs. \cite{pfeifer1993,pfeifer1995}, in which he proposes that general time-energy uncertainty relations for time-dependent Hamiltonians should be computable without solving Schr\"{o}dinger's equation. The main result reads as follows: given a quantum state $\Ket{\psi(t)}$ which evolves according to $i\frac{d}{dt}\Ket{\psi(t)}=H(t)\Ket{\psi(t)}$ with $\Ket{\psi(0)}=\Ket{\psi_0}$, and an arbitrary reference state $\Ket{\varphi}$, then the following relation holds
\begin{equation}
\lvert\BraKet{\varphi}{\psi(t)}\rvert  \lesseqgtr \mathrm{sin}_*\left(\delta\pm h(t)\right),
\label{ec:qsl_pfei1}
\end{equation}

\noindent where $\delta=\mathrm{arcsin}(\lvert\BraKet{\varphi}{\psi_0}\rvert)=\frac{\pi}{2}-\mathrm{arccos}(\lvert\BraKet{\varphi}{\psi_0}\rvert)$, sin$_*$ is the a modified sine function
\begin{equation}
\mathrm{sin}_*(x)=\left\{\begin{array}{rcl}
0 & \mathrm{if} & x\leq0 \\
\mathrm{sin}(x) & \mathrm{if} & 0<x\leq 1\\
1 & \mathrm{if} & x>1 \end{array}\right.
\label{ec:qsl_pfei_sin}
\end{equation}
\noindent and we defined
\begin{equation}
h(t)=\min\limits_{\Ket{\psi_0},\Ket{\varphi}}\left\lbrace\int_0^t \Delta E_{\varphi} (t') \:dt',\int_0^t \Delta E_{\psi_0} (t')\:dt'\right\rbrace,
\label{ec:qsl_pfei_h}
\end{equation}
\noindent where we used the notation $\Delta E_{\chi}\equiv\Bra{\chi}H^2\Ket{\chi}-\Bra{\chi}H\Ket{\chi}^2$. Pfeifer's relation (\ref{ec:qsl_pfei1}) is appealing to the quantum control problem studied here, since it gives bounds for the probability of finding a \emph{driven} system in an \emph{arbitrary} state $\Ket{\varphi}$ \cite{pfeifer1993}. More interestingly, we can extract a bound on the evolution time itself, in the following way. If we consider the upper bound in (\ref{ec:qsl_pfei1}) for such probability, and consider the reference state to be our target state, $\Ket{\varphi}=\Ket{\psi_g}$, we get that, at time $t=T$
\begin{equation}
\lvert\BraKet{\psi_g}{\psi(T)}\rvert  \leq \mathrm{sin}_*\left(\delta+ h(T)\right).
\label{ec:qsl_pfei2}
\end{equation}
From this expression its clear that, in order to have a successful control process, we need the upper bound to be as large as possible, i.e. 1. Looking at the definition (\ref{ec:qsl_pfei_sin}), it is then sufficient to impose
\begin{equation}
\delta + h(T)\geq \frac{\pi}{2}\Rightarrow h(T)\geq\frac{\pi}{2}-\delta=\frac{1}{2}s(\psi_0,\psi_g).
\label{ec:qsl_pfei3}
\end{equation}

Note that $h(T)$ depends on $T$ via the control field $\vec{u}(T)$. In order to obtain a lower bound for the evolution time, we proceed as we did when deriving (\ref{ec:qsl_new_brody2}) and bound the integral in (\ref{ec:qsl_pfei_h}) by
\begin{equation}
h(T)\leq \Delta E_{\chi}^{max} T\ \mathrm{with}\ \chi=\psi_0,\psi_g,
\end{equation}
\noindent where, again, we expect $\Delta E_{\chi}^{max}$ to be an explicit function of $\{u_i^{max}\}$. Rearranging the expression above we arrive at 

\begin{equation}
T\geq\frac{s(\psi_0,\psi_g)}{2\Delta E_{\chi}^{max}}\equiv t_{min}^{B}\ \mathrm{with}\ \chi=\psi_0\:\mathrm{or}\:\psi_g.
\label{ec:qsl_pfeifer}
\end{equation}

Again, $t_{min}^B$ is also of the form (\ref{ec:qsl_new_gen}) and thus allows us to obtain a lower bound on the minimum evolution time without knowing the actual shape of $\vec{u}(t)$.\\

We now explore an interesting property of Pfeifer's bound (\ref{ec:qsl_pfeifer}). Assume the Hamiltonian of the system has the form
\begin{equation}
H(u(t))=H_0+u(t)H_c,
\label{ec:qsl_hamilt_lineal}
\end{equation}
\noindent where we suppose that the control field $u(t)$ has dimensionless units. We can then explicitly write down the variance of the Hamiltonian as 

\begin{equation}
\Delta E^2=\Delta H_0^2 + u^2\Delta H_c^2 + u(\langle\lbrace H_0,H_c\rbrace\rangle-2\langle H_0\rangle \langle H_c\rangle)
\label{ec:qsl_varian}
\end{equation}

Suppose now that our control problem is such that the initial and target states $\Ket{\psi_0}$, $\Ket{\psi_g}$ are eigenstates of $H_c$. Then, we trivially obtain that $\Delta H_c=0$, but also that the crossed term in (\ref{ec:qsl_varian}) vanishes. Inserting this into expression (\ref{ec:qsl_pfeifer}) we get

\begin{equation}
t_{min}^B=\frac{s(\psi_0,\psi_g)}{\mathrm{min}\lbrace \Delta H_0\lvert_{\psi_0},\Delta H_0\lvert_{\psi_g}\rbrace}.
\end{equation}
What is interesting about this result is that it is completely independent of $u(t)$; not only of its actual temporal shape, but also of its maximum possible value. This means that, even in an unconstrained control problem where $u^{max}\rightarrow\infty$, there is still a fundamental limit for the speed in which we can control the system. That limit is set only by the initial and final states, and the free Hamiltonian $H_0$. Note than analogous bound can be found if $\Ket{\psi_0}$, $\Ket{\psi_g}$ are eigenstates of $H_0$. \\

%

Finally, we present a third method for obtaining a bound of the form (\ref{ec:qsl_new_gen}). We begin by considering two arbitrary time-dependent Hamiltonians $H_1$ and $H_2$, and two respective states $\Ket{\psi_1(t)}$ and $\Ket{\psi_2(t)}$ such that $\frac{d}{dt}\Ket{\psi_k}=-i H_k(t)\Ket{\psi_k(t)}$ with $k=1,2$ and $\Ket{\psi_1(0)}=\Ket{\psi_2(0)}=\Ket{\psi_0}$. We can then write
\begin{equation}
\frac{d}{dt}\BraKet{\psi_1}{\psi_2}=i\Bra{\psi_1}(H_1-H_2)\Ket{\psi_2},
\end{equation}

\noindent and then integrate the above expression from $t=0$ to $t=T$, which yields
\begin{widetext}
	
\begin{eqnarray}
\BraKet{\psi_1(T)}{\psi_2(T)}-1 &=&i\int_0^T
\Bra{\psi_1(t')}(H_1(t')-H_2(t'))\Ket{\psi_2(t')}\:dt' \nonumber \\
\Rightarrow \lvert\BraKet{\psi_1(T)}{\psi_2(T)}-1\rvert&\leq&\int_0^T \lvert\Bra{\psi_1(t')}(H_1(t')-H_2(t'))\Ket{\psi_2(t')}\rvert\:dt' \leq \int_0^T \|H_1(t')-H_2(t')\|\:dt'. \label{ec:qsl_rabitz1}
\end{eqnarray}

\end{widetext}

We now take an approach proposed by Arenz \etal \cite{arenz2017}. We consider $H_1$ to be of the form (\ref{ec:qsl_hamilt_lineal}), i.e. $H_1=H_0+u(t)H_c$, and also fix $H_2=u(t)H_c$. For a successful control protocol, we have that $\Ket{\psi_1(T)}=\Ket{\psi_g}$, and we can also integrate $\Ket{\psi_2(t)}$ up to $t=T$, which trivially yields $\Ket{\psi_2(T)}=\expo{-i \alpha(T) H_c}\Ket{\psi_0}$ where $\alpha(T)=\int_0^T u(t')dt'$. In this case, expression (\ref{ec:qsl_rabitz1}) can be casted as
\begin{equation}
|\Bra{\psi_g}\ex{-i\alpha(T)H_c}\Ket{\psi_0}|\leq \|H_0\|\:T.
\end{equation}

We can further bound this expression in order to get rid of the dependence on the unknown function $u(t)$. To do so, we use the spectral decomposition of $H_c=\sum_j \eps_j^c \KetBra{\phi_j^c}{\phi_j^c}$ and the inequality $|\sum_j z_j -1|\geq 1-\sum_j |z_j|$ (with $|z_j|\leq 1$) to obtain
\begin{equation}
1-\sum_j^n \lvert \BraKet{\psi_g}{\phi_j^c}\BraKet{\phi_j^c}{\psi_0}|\leq \|H_0\|T,
\end{equation}
\noindent which then gives us a new bound of the desired form (\ref{ec:qsl_new_gen})
\begin{equation}
T\geq \frac{1-\sum\limits_j^n\lvert \BraKet{\psi_g}{\phi_j^c}\BraKet{\phi_j^c}{\psi_0}\rvert}{\|H_0\|}\equiv t_{min}^{C1}.
\label{ec:qsl_r1}
\end{equation}

A similar expression can be derived in an analogous fashion by choosing $H_2=H_0$. In that case we obtain
\begin{equation}
T\geq \frac{1-\sum\limits_j^n\lvert \BraKet{\psi_g}{\phi_j^0}\BraKet{\phi_j^0}{\psi_0}\rvert}{u^{max}\|H_c\|}\equiv t_{min}^{C2},
\label{ec:qsl_r2}
\end{equation}
\noindent where now $\lbrace \Ket{\phi_j^0}\rbrace$ are eigenstates of the free Hamiltonian $H_0$. Expressions (\ref{ec:qsl_r1}) and (\ref{ec:qsl_r2}) provide different ways to bound evolution times in quantum control problems. An interesting feature of these is that they are \emph{explicit} functions of $\psi_0$, $\psi_g$, $H$ and $u^{max}$, as opposed to the two previous results (\ref{ec:qsl_new_brody}) and (\ref{ec:qsl_pfeifer}), where the actual dependence on $H$ and $u^{max}$ has to be worked out on each particular problem. This means that, for example, $t_{min}^{C1}$ will always give a result independent of $u^{max}$ regardless the initial and target states.\\

\section{Application to a two-level system}
\label{sec:two_level}

In the previous section we analyzed an approach for bounding evolution times in driven quantum systems, which differs from the standard QSL. The goal was to obtain as much information as possible about the evolution time without needing to solve the dynamics of the system. In this section we will apply these results to the example of a driven two-level system. For this we consider the following Hamiltonian,
\begin{equation}
H(u)=\left( 
\begin{array}{c c}
u & \frac{\Delta}{2} \\
\frac{\Delta}{2} & -u
\end{array} \right)= u\:\sigma_z+\frac{\Delta}{2}\sigma_x,
\label{ec:qoc_lz_hami}
\end{equation}

\noindent where $\sigma_i$, $i=x,y,z$ is a Pauli operator, $\Delta$ is a parameter that we consider fixed and $u$ is the control parameter. We define $\Ket{g_\gamma}$ to be the ground state of $H(\gamma)$ (i.e. its eigenstate with negative eigenvalue). We focus on the following control problem: we start in the initial state $\Ket{\psi_0}=\Ket{g_{-\gamma}}$ and we wish to drive the system to the target state $\Ket{\psi_g}=\Ket{g_{+\gamma}}$ (here $\gamma>0$). Moreover, we wish to do so in the minimum possible time. The problem of finding the required control field for this process was solved by Hegerfeldt \cite{hegerfeldt2013}, who proved that different protocols arise whether we place constraints on the amplitude $|u(t)|$ of the control field or not. In the unconstrained case, the optimal field is
\begin{equation}
u(t)=\left\{\begin{array}{lcr}
+u_0 & \mathrm{for} & 0<t<t_0 \\
0 & \mathrm{for} & t_0<t<t_0+T \\
-u_0 & \mathrm{for} & t_0+T<t<2t_0+T \end{array}\right.,
\label{ec:qoc_comp} 
\end{equation}
\noindent where $u_0\gg \Delta$, $u_0t_0=\pi/4$, and as $|u(t)|$ has no restrictions, we can choose $u_0\rightarrow\infty$ so as to have $t_0\rightarrow0$. The total evolution time is then given by
\begin{equation}
T_{opt}^{(1)}=T + 2t_0\rightarrow T=\frac{2}{\Delta}\mathrm{arctan}\left(\frac{2\gamma}{\Delta}\right)=\frac{\pi -  2 \theta}{\Delta},
\label{ec:qoc_tmin1}
\end{equation}

\noindent where we have introduced the angle $\theta$ as an alternative parametrization of the initial state, $\tan(\theta)=\frac{\Delta}{2\gamma}$. In the constrained case, where $|u(t)|\leq u_{max} \equiv \Lambda$, the optimal solution is similar, 
\begin{equation}
u(t)=\left\{\begin{array}{lcr}
+\Lambda & \mathrm{for} & 0<t<T_\Lambda \\
0 & \mathrm{for} & T_\Lambda<t<T_\Lambda+T_{off} \\
-\Lambda & \mathrm{for} & T_\Lambda+T_{off}<t<2T_\Lambda+T_{off} \end{array}\right.
\label{ec:qoc_const} 
\end{equation}
The evolution time here is given by 
\begin{equation}
T_{opt}^{(2)}=T_{off}+2T_\Lambda.
\label{ec:qoc_tmin2}
\end{equation}
The optimal values of $T_\Lambda$ and $T_{off}$ differ whether the maximum field $\Lambda$ is smaller or larger than ${\Delta^2}/(4\gamma)$. The corresponding expressions are a bit cumbersome and are given in the Appendix.\\

Here we will be interested in comparing the actual optimal control times of eqn. (\ref{ec:qoc_tmin1}) and (\ref{ec:qoc_tmin2}) with the bounds given in the previous section. Again we emphasize that, in order to evaluate  the QSL time $T_{QSL}^*$, c.f. eqn. (\ref{ec:qsl_deff2}), we would need to know how the system evolves under the optimal protocol. For each case (i.e. constrained or unconstrained), $T_{QSL}^*(\theta)$ can be worked out, as was done in \cite{poggi2013}. We give the corresponding expressions in the Appendix as well. \\

We now turn to computing the new bounds $t_{min}^X$ with X=A, B, C1 and C2, which are of the form
\begin{equation}
T\geq t_{min}\left(H,u_{max},\Ket{\psi_0},\Ket{\psi_g}\right).
\end{equation}
We stress that, since these expressions are independent of the actual dynamics of the system, we will derive them for the constrained and unconstrained protocols in the same way. This is a key aspect of the approach we propose, since we should be able to obtain some information about the minimum evolution time without any knowledge about the actual optimal protocol. Let us start with $t_{min}^A$ of eqn (\ref{ec:qsl_brody}), for which we calculate the norm of $H$
\begin{equation}
\|H\|=\sqrt{\mathrm{tr}\left(H^2\right)}=\sqrt{2\left(\frac{\Delta^2}{4}+u^2\right)}
\end{equation}
We bound this expression to obtain
\begin{equation}
t_{min}^A=\frac{\frac{\pi}{2}-\theta}{\sqrt{\frac{\Delta^2}{4}+u_{max}^2}}
\label{ec:qoc_qsl_b}
\end{equation}

For computing the bound (\ref{ec:qsl_pfeifer}) obtained via Pfeifer's theorem, $t_{min}^B$, we need to evaluate the variance $\Delta E$ of $H$ in both the initial and final states. This can be done in a straightforward way, and we obtain
\begin{equation}
\Delta E\rvert_{\psi_0/\psi_g}=\frac{\Delta}{2}\cose{\theta}\lvert 1\pm\frac{2u}{\Delta}\mathrm{tan}\left(\theta\right)\rvert,
\end{equation}
\noindent which in turn gives
\begin{equation}
h(t)=\frac{\Delta}{2}\cose{\theta}\int_0^t\mathrm{min}\left\{\lvert 1\pm\frac{2u(t)}{\Delta}\mathrm{tan}\left(\theta\right)\rvert\right\}
\end{equation}
In this way we obtain
\begin{equation}
t_{min}^B=\frac{\frac{\pi}{2}-\theta}{\frac{\Delta}{2}\cose{\theta}+u_{max}\seno{\theta}}
\label{ec:qoc_qsl_p}
\end{equation}

We finally consider $t_{min}^{C1}$, which was defined in eqn. (\ref{ec:qsl_r1}).
We recall that here $H_0=\frac{\Delta}{2}\sigma_x$ is the free term of the Hamiltonian, and $\Ket{\phi_j^c}$ refer to $\Ket{0}$ and $\Ket{1}$, i.e. the eigenstates of the control operator $\sigma_z$. Straightforward calculation gives
\begin{equation}
t_{min}^{C1}=\frac{1-\seno{\theta}}{\frac{\sqrt{2}}{2}\Delta}.
\label{ec:qoc_qsl_r1}
\end{equation}
We point out that $t_{min}^{C2}$ defined in eqn. (\ref{ec:qsl_r2}) turns out to be 0 for this problem, for all values of $\theta$.\\ 

Up to this point we have computed three bounds for the evolution time in this control problem (\ref{ec:qoc_qsl_b}), (\ref{ec:qoc_qsl_p}) and (\ref{ec:qoc_qsl_r1}) which are computed without knowledge of the solution to the time-optimal control problem. We also have, from \cite{poggi2013}, the corresponding QSL time for as a function of $\theta$, $T_{QSL}^*(\theta)$ (see Appendix for the explicit expressions) which is computed using such time-optimal solution. Let us first compare all of these expressions with the optimal time $T_{opt}$ for the case of full population transfer, i.e. $\gamma\rightarrow\infty$ or $\theta\rightarrow 0$. In this case, $T_{opt}=\frac{\pi}{\Delta}$, while
\begin{equation}
T_{QSL}^*=t_{min}^B=\frac{\pi}{\Delta}>\frac{\sqrt{2}}{\Delta}=t_{min}^{C1}.
\end{equation}

Since these were the geometrical expressions, it is reasonable to have obtained a tight bound: when $\theta=0$, the optimal evolution (which is generated by setting $u=0$) is along a geodesic, which is precisely when the Anandan-Aharanov relation is saturated. For the remaining expression, we obtain $t_{min}^A=0$ due to the dependence on $u_{max}\rightarrow\infty$. It is interesting to see that Pfeifer's bound $t_{min}^B$ matches the optimal evolution time also, although we didn't use any information about the optimal solution itself to compute it. This result gives us confidence about the usefulness of this method to bound evolution times in optimal control problems.\\

\begin{figure}[t!]
	\centering
	\includegraphics[width=0.8\linewidth]{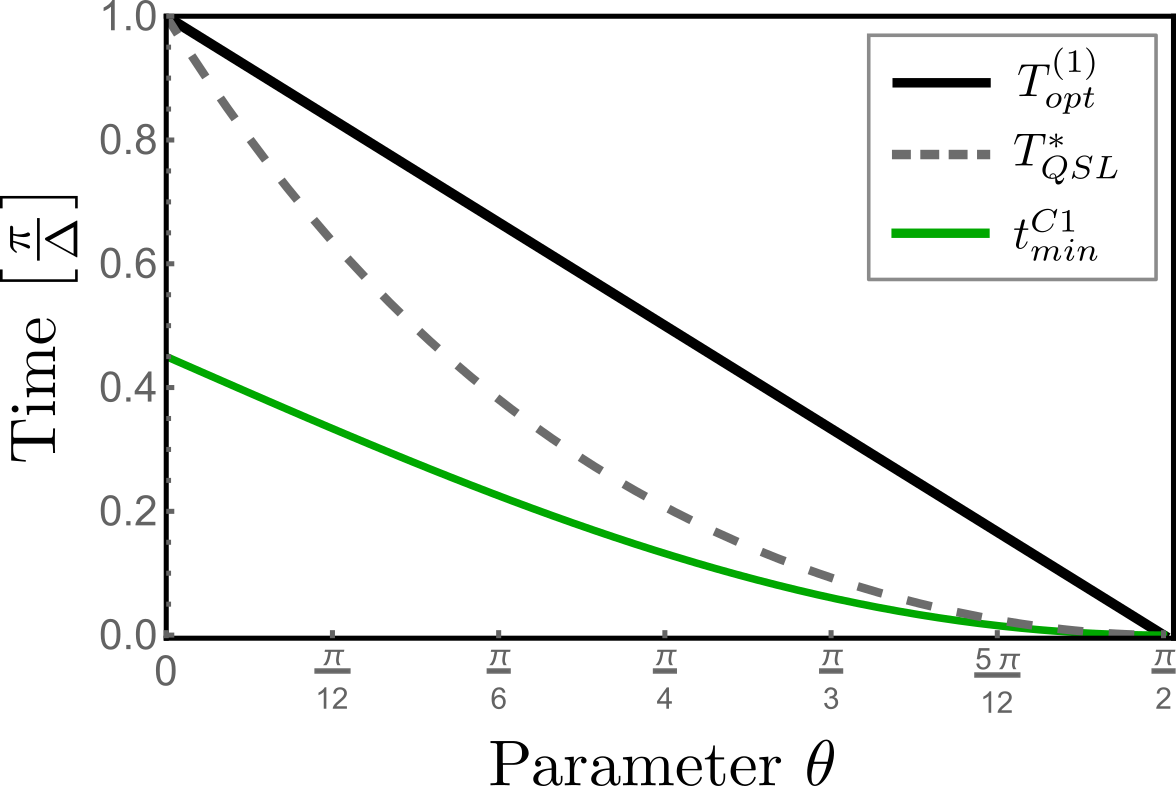}
	\caption{Optimal evolution time $T_{opt}$, together with QSL time $T_{QSL}^*$ and bound $t_{min}^{C1}$ obtained from eqn.  (\ref{ec:qoc_qsl_r1}) for the composite-pulse protocol (with unconstrained $u(t)$) as a function of parameter $\theta$.}
	\label{fig:qsl_unconst}
\end{figure}

Let us now analyze the general case of finite $\gamma$. For unconstrained control, we have that $u_{max}\rightarrow\infty$. Note that this immediately gives $t_{min}^A=t_{min}^B=0$ (recall also that $t_{min}^{C2}=0$), but $t_{min}^{C1}$ remains nonzero since it does not depend on the control field constraints, as we pointed out in the previous section. In Fig. \ref{fig:qsl_unconst} we plot this quantity along with the actual optimum time $T_{opt}$ as a function of angle $\theta$, which defines the initial and target states. Note that for $\theta=\pi/2$ ($\gamma=0$) both states are the same, and thus $T_{opt}=0$. Note also that $t_{min}^{C1}$, which was computed without knowledge of the optimal evolution, is never tight (except for $\theta=\frac{\pi}{2}$, which is trivial). However, its interesting to point out that it is nonzero in spite of the fact that the control field is unconstrained (and is infinite in this case), and thus gives a meaningful bound as opposed to $t_{min}^A$ and $t_{min}^B$.\\

\begin{figure}[t!]
	\centering
	\includegraphics[width=0.8\linewidth]{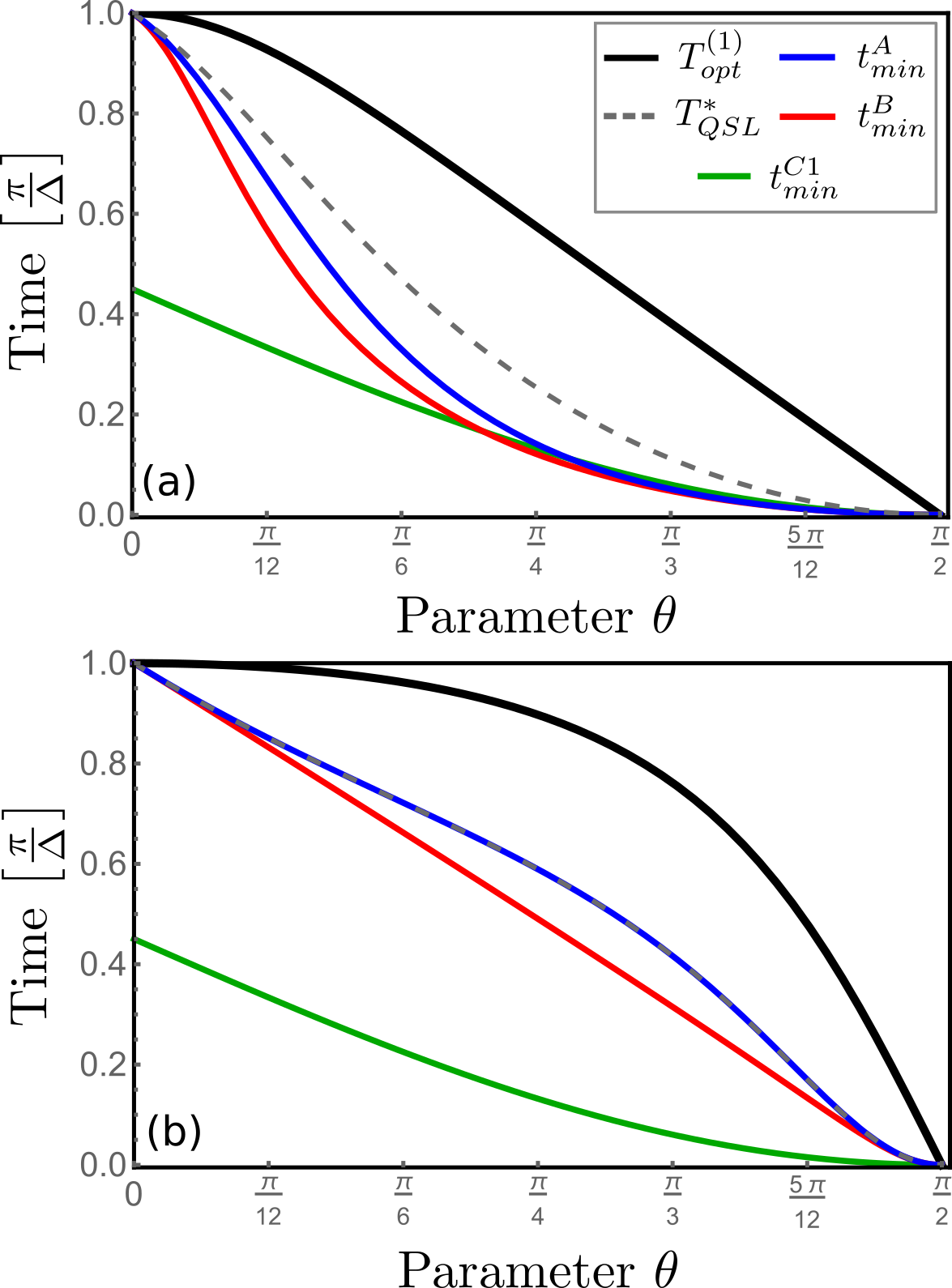}
	\caption{Optimal evolution time $T_{opt}$, together with QSL time $T_{QSL}^*$ and its bounds obtained from the expressions discussed in the text for: (a) $\Lambda>\frac{\Delta^2}{4\gamma}$ (in this calculations $\Lambda=6\frac{\Delta^2}{4\gamma}$) and (b) $\Lambda\leq\frac{\Delta^2}{4\gamma}$ (in this calculations $\Lambda=0.2\frac{\Delta^2}{4\gamma}$). Note that in this last case, $T_{QSL}^*=t_{min}^B$.}
	\label{fig:qsl_const}
\end{figure}

We now compare the bounds for the case of constrained control, where $|u(t)|\leq \Lambda$. As already mentioned, here the optimal solution depends on the relation between $\Lambda$ and $\gamma$. For $\Lambda \geq \frac{\Delta^2}{4\gamma}$, we have the bang-off-bang protocol described by expressions (\ref{ec:qoc_const}) and (\ref{ec:qoc_tbob}),while for $\Lambda < \frac{\Delta^2}{4\gamma}$, the solution is the bang-bang protocol, c.f. eqn. (\ref{ec:qoc_const}) and (\ref{ec:qoc_tbb}). In Fig. \ref{fig:qsl_const} (a) we show results for the bang-off-bang case. All the bounds considered yield different curves in general. Moreover, there is no bound tighter than another for all $\theta$. Of all the bounds computed without the optimal protocol, $t_{min}^B$ stands out as the better one. In Fig. \ref{fig:qsl_const} (b) we show results for the bang-bang case. Interestingly, in this case $\Delta E$ is constant throughout the evolution, albeit the Hamiltonian being time-dependent itself. As a result, $t_{min}^B$ is equal to the Mandelstam-Tamm bound from the time-independent case, and is tighter than $t_{min}^{C1}$ as before. We thus find that the bound derived from Pfeifer's theorem $t_{min}^B$ is bigger or equal than all of the others for all $\theta$, and results in the tighter
bound, albeit being computed without knowledge of the optimal protocol. This result provides further evidence about the usefulness of this particular expression for bounding minimal evolution times in quantum control problems.\\

\section{Outlook and final remarks}
\label{sec:final}

In this paper we have revisited the quantum speed limit (QSL) formulation for unitary dynamics driven by time-dependent Hamiltonians, focusing on its application to quantum control problems. We argued that the QSL is not usually useful to obtain lower bounds on control times before solving the optimal control problem. The reason behind this is that the QSL time depends implicitly on the actual evolution of the system, which is a priori unknown apart from the initial and final (target) state. However, obtaining such bounds is interesting and could actually be helpful to tackle the optimization, since in principle it would allow one to rule out all possible control times lower than the bound. With this in mind, here we have proposed a number of properties that a lower bound should have in order to be useful for control applications, c.f. eqn. (\ref{ec:qsl_new_gen}). The main such property is that the bound should be computable without knowing the full time-dependent state. Then we have put together (and in some cases adapted and further developed), previous results related to optimal control and QSL that actually have this properties. We studied these new lower bounds on control times for a two-level system, for which the time-optimal control problem has been analytically solved. We found that in all cases this new formulation gives meaningful bounds, and provides information which is comparable to the one obtained with the standard QSL, albeit being calculated without knowing the optimal control solution. We point out that the ideas layed down here for new bounds on control times could in principle be extended to open quantum systems, using the approach in Pfeifer's theorem (\ref{ec:qsl_pfei1}) applied to a metric like the relative purity between states. More generall,y while these results are encouraging, it is expected that the proposed bounds will not scale favorably with system size \cite{arenz2017}, as happens with the geometric QSL itself \cite{bukov2019}. As a consequence, further work is needed to find new techniques to bound control times for quantum systems, but we believe that such techniques could benefit from the results presented in this work.

\section*{Acknowledgments}

The author gratefully acknowledges Fernando Lombardo and Diego Wisniacki for their continued support as advisors. This work received supported by CONICET, UBACyT, ANPCyT (Argentina) and National Science Foundation (NSF) grant no. PHY-1630114 (USA).

\bibliography{paper_afa} 

\appendix

\section{Optimal control times for the constrained problem}

Here we give the explicit form of the times $T_{\Lambda}$ and $T_{off}$ derived by Hegerfeldt in \cite{hegerfeldt2013}. For $\Lambda\geq\frac{\Delta^2}{4\gamma}$, we have
\begin{eqnarray}
T_\Lambda&=&\frac{1}{\sqrt{\Lambda^2+\frac{\Delta^2}{4}}}\mathrm{arcsin}\left(\sqrt{\frac{\Lambda^2+\frac{\Delta^2}{4}}{2\Lambda(\Lambda+\gamma)}}\right) \nonumber \\
T_{off}&=&\frac{2}{\Delta}\mathrm{arctan}\left(\frac{\Lambda\gamma-\frac{\Delta^2}{4}}{\frac{\Delta}{2}\sqrt{\Lambda^2+2\Lambda\gamma-\frac{\Delta^2}{4}}}\right),
\label{ec:qoc_tbob}
\end{eqnarray}
\noindent which is called a 'bang-off-bang' protocol, while for $\Lambda<\frac{\Delta^2}{4\gamma}$, the result is
\begin{eqnarray}
T_\Lambda&=&\frac{1}{\sqrt{\Lambda^2+\frac{\Delta^2}{4}}}\mathrm{arcsin}\left(\sqrt{\frac{\gamma\left(\Lambda^2+\frac{\Delta^2}{4}\right)}{\frac{\Delta^2}{2}(\Lambda+\gamma)}}\right) \nonumber \\
T_{off}&=&0
\label{ec:qoc_tbb}
\end{eqnarray}

\noindent which is typically termed 'bang-bang'.\\

Also, we give expressions for the QSL time for both cases of interest. All of these results were obtained in \cite{poggi2013} and so don't derive them again here. For the unconstrained problem ($u_{max}=\infty$), we have that

\begin{equation}
T_{QSL}^*(\theta) = \frac{s(\theta)T_{opt}^{(1)}(\theta)}{s(\theta) + \pi \sin(\theta)},
\end{equation}

\noindent where we defined $s(\theta)=\pi - 2\theta$. For the constrained problem ($u_{max}<\infty$), for the bang-off-bang protocol we have

\begin{equation}
T_{QSL}^*(\theta) = \frac{s(\theta)T_{opt}^{(2)}(\theta)}{4\left(\Lambda \sin(\theta) + \frac{\Delta}{2} \cos(\theta)\right)T_{\Lambda}(\theta) + \Delta T_{off}(\theta)}
\end{equation}

\noindent while for the bang-bang protocol the QSL time is

\begin{equation}
T_{QSL}^*(\theta) = \frac{s(\theta)}{2\left(\Lambda \sin(\theta) + \frac{\Delta}{2} \cos(\theta)\right)}.
\end{equation}

\end{document}